\documentclass[final            
    ,numberedheadings 
]{aipproc}

\layoutstyle{8x11single}

\usepackage{amsfonts}
\usepackage{amsmath}
\usepackage{bm}
\usepackage{amsopn}

\newcounter{theorem}  \newenvironment{theorem}
{\refstepcounter{theorem}
   \vspace{1 em}
   \noindent{\bf Theorem~\thetheorem}
   \begin{em}}
  {\end{em}
   \newline}

\allowdisplaybreaks

\begin{document}

\title{Post-Newtonian Theory and Dimensional Regularization}


\author{Luc Blanchet}{
  address={${\mathcal{G}}{\mathbb{R}}
\varepsilon{\mathbb{C}}{\mathcal{O}}$, Institut d'Astrophysique de
Paris, UMR 7095-CNRS, 98$^{\text{bis}}$ boulevard Arago, 75014 Paris,
France}}

\begin{abstract}
Inspiralling compact binaries are ideally suited for application of a
high-order post-Newtonian (PN) gravitational wave generation formalism.
To be observed by the LIGO and VIRGO detectors, these very relativistic
systems (with orbital velocities $v\sim 0.5c$ in the last rotations)
require high-accuracy templates predicted by general relativity theory.
Recent calculations of the motion and gravitational radiation of compact
binaries at the 3PN approximation using the Hadamard self-field
regularization have left undetermined a few dimensionless coefficients
called ambiguity parameters. In this article we review the application
of dimensional self-field regularization, within Einstein's classical
general relativity formulated in $D$ space-time dimensions, which
finally succeeded in clearing up the problem, by uniquely fixing the
values of all the ambiguity parameters.
\end{abstract}

\maketitle


\section{Introduction}\label{secI}
The problem of the motion and gravitational radiation of compact objects
in post-Newtonian (PN) approximations of general relativity is of
crucial importance, for at least three reasons. First, the motion of $N$
objects at the 1PN level,\footnote{As usual $n$PN refers to the terms of
order $(v/c)^{2n}$ where $v$ is the typical orbital velocity of the
objects and $c$ is the speed of light. We shall generally abbreviate
$(v/c)^{2n}$ as $1/c^{2n}$.} according to the Einstein--Infeld--Hoffmann
equations~\cite{EIH}, is routinely taken into account to describe the
Solar System dynamics. Second, the gravitational radiation-reaction
force, in reaction to the emission of gravitational radiation, which
appears in the equations of motion at the 2.5PN order $\sim 1/c^{5}$,
has been experimentally verified, by the observation of the secular
acceleration of the orbital motion of the Hulse--Taylor binary pulsar
PSR~1913+16~\cite{TFMc79, TW82, DT91, T93}.

Last but not least, the forthcoming detection and analysis of the
gravitational waves emitted by inspiralling compact binaries --- two
neutron stars or black holes driven into coalescence by emission of
gravitational radiation --- will necessitate the prior knowledge of the
equations of motion and radiation field up to high post-Newtonian order.
Inspiralling compact binaries are extremely promising sources of
gravitational waves for the detectors LIGO, VIRGO, GEO and TAMA. The two
compact objects steadily lose by gravitational radiation their orbital
binding energy; as a result, the orbital separation between them
decreases, and the orbital frequency increases. The frequency of the
gravitational-wave signal, which equals twice the orbital frequency for
the dominant harmonics, ``chirps'' in time (\textit{i.e.} the signal
becomes higher and higher pitched) until the two objects collide and
merge.

Strategies to detect and analyze the very weak signals from compact
binary inspiral involve matched filtering of a set of accurate
theoretical template waveforms against the output of the detectors.
Several analyses~\cite{3mn, CF94, P95, DIS98, BCV03a, BCV03b} have shown
that, in order to get sufficiently accurate theoretical templates, one
must include post-Newtonian effects up to the 3PN level at least. To
date, the templates have been completed through 3.5PN order for the
phase evolution~\cite{BDIWW95, BFIJ02, BDEI04}, and 2.5PN order for the
amplitude corrections~\cite{BIWW96, ABIQ04}. Spin effects are known for
the dominant relativistic spin-orbit coupling term at 1.5PN order and
the spin-spin coupling term at 2PN order~\cite{KWW93, ACST94, K95,
GerPV98, Ger00a, Ger00b}, and also for the next-to-leading spin-orbit
coupling at 2.5PN order~\cite{OTO98, TOO01, FBB06spin, BBF06spin}.

The main point about modelling the inspiralling compact binary is that a
model made of two \textit{structureless point particles}, characterized
solely by two mass parameters $m_1$ and $m_2$ (and possibly two spins),
is sufficient. Indeed, most of the non-gravitational effects usually
plaguing the dynamics of binary star systems, such as the effects of a
magnetic field, of an interstellar medium, and so on, are dominated by
gravitational effects. However, the real justification for a model of
point particles is that the effects due to the finite size of the
compact bodies are small. Consider for instance the influence of the
Newtonian quadrupole moments $\mathrm{Q}_1$ and $\mathrm{Q}_2$ induced
by tidal interaction between two neutron stars. We suppose that the
neutron stars have no intrinsic spins. Let $a_1$ and $a_2$ be the radius
of the stars, and $L$ the distance between the two centers of mass. We
have, for tidal moments,
\begin{equation}
  \mathrm{Q}_1 = k_1 m_2 \frac{a_1^5}{L^3}, \qquad \mathrm{Q}_2 = k_2 m_1
  \frac{a_2^5}{L^3}, \label{Q}
\end{equation}
where $k_1$ and $k_2$ are the star's dimensionless (second) Love
numbers, which depend on their internal structure, and are, typically,
of the order unity. On the other hand, for compact objects, we can
introduce their ``compactness'', defined by the dimensionless ratios
\begin{equation}
  K_1 = \frac{G m_1}{a_1 c^2},\qquad K_2 = \frac{G m_2}{a_2 c^2},
  \label{K}
\end{equation}
which equal $\sim 0.2$ for neutron stars (depending on their equation of
state). The quadrupoles $\mathrm{Q}_1$ and $\mathrm{Q}_2$ will affect
both the Newtonian binding energy $\mathcal{E}$ of the two bodies, and
the emitted total gravitational wave flux $\mathcal{F}$ as computed,
say, using the standard Einstein quadrupole formula. It is known that
for inspiralling compact binaries the neutron stars are not co-rotating
because the tidal synchronization time is much larger than the time left
till the coalescence. The best models for the fluid motion inside the
two neutron stars are the so-called Roche--Riemann
ellipsoids~\cite{Kochanek}, which have tidally locked figures (the
quadrupole moments face each other at any instant during the inspiral),
but for which the fluid motion has zero circulation in the inertial
frame. In the Newtonian approximation we find that within such a model
(in the case of two identical neutron stars) the orbital phase, $\phi =
\int\omega dt$ where $\omega$ is the orbital frequency, reads
\begin{equation}
  \phi - \phi_0 = \frac{1}{8x^{5/2}} \left\{1+\mathrm{const\ } k
  \left(\frac{x}{K}\right)^5\right\}, \label{8}
\end{equation}
where $x=(G m\omega/c^3)^{2/3}$ is a standard dimensionless PN parameter
$\sim 1/c^2$, and where $k$ is the Love number and $K$ is the
compactness of the neutron star. The first term in the right-hand-side
(RHS) of~(\ref{8}) corresponds to the gravitational-wave damping of two
point masses; the second term describes the finite-size effect, which
appears as a relative correction, proportional to $(x/K)^5$, to the
latter radiation damping effect. Because the finite-size effect is
purely Newtonian, its relative correction $\sim (x/K)^5$ should not
depend on $c$; and indeed the factors $1/c^2$ cancel out in the ratio
$x/K$. However, the compactness $K$ of compact objects is by
Eq.~(\ref{K}) of the order unity (or, say, a few tenths), therefore the
$1/c^2$ it contains should not be taken into account in order to find
the magnitude of the effect in this case, and so the real order of
magnitude of the relative contribution of the finite-size effect in
Eq.~(\ref{8}) should be given by $x^5$ alone. This means that for
non-spinning compact objects the finite-size effect should be
comparable, numerically, to a post-Newtonian correction of magnitude
$x^5 \sim 1/c^{10}$ namely 5PN order.\,\footnote{This result can be
derived in the context of relativistic equations of motion, and yields a
proof of the so-called ``effacement'' principle in general relativity,
according to which the internal structure of the compact bodies does not
show up in their motion and emitted radiation which depend only on the
masses~\cite{D83houches}.} This is a much higher post-Newtonian order
than the one at which we shall investigate the gravitational effects on
the phasing formula. Using $k'\equiv (\mathrm{const} \ k)\sim 1$ and
$K\sim 0.2$ for neutron stars (and the bandwidth of a VIRGO detector
between 10~Hz and 1000~Hz), we find that the cumulative phase error due
to the finite-size effect amounts to less that one orbital rotation over
a total of $\sim 16\,000$ produced by the gravitational-wave damping of
point masses. The conclusion is that the finite-size effect can in
general be neglected in comparison with purely gravitational-wave
damping effects.\,\footnote{But note that for non-compact or moderately
compact objects (such as white dwarfs) the Newtonian tidal interaction
dominates over the radiation damping.} Thus the appropriate theoretical
description of inspiralling compact binaries is by two point masses
within the post-Newtonian approximation.

Our strategy to obtain the motion and radiation of a system of two
point-like particles at the 3PN order is to start with a general form of
the 3PN metric, that is valid for a general continuous (smooth) matter
distribution. Applying such metric to a system of point particles, we
find that most of the integrals become divergent at the location of the
particles, \textit{i.e.} when $\mathbf{x}\rightarrow\mathbf{y}_1(t)$ or
$\mathbf{y}_2(t)$, where $\mathbf{y}_1(t)$ and $\mathbf{y}_2(t)$ denote
the two trajectories. Consequently, we must supplement the calculation
by a prescription for how to remove the ``infinite part'' of these
integrals. At this stage different choices for a ``self-field''
regularization, which will take care of the infinite self-field of point
particles, are possible. Among them:
\begin{enumerate}
\item Hadamard's self-field regularization, which has proved to be very
  convenient for doing practical computations (in particular, by
  computer), but suffers from the important drawback of yielding some
  ambiguity parameters, which cannot be determined within this
  regularization, at the 3PN order;
\item Dimensional self-field regularization, an
  extremely powerful regularization, that is free of any ambiguities (at
  least up to the 3PN level), and as we shall see permits to uniquely
  fix the values of the ambiguity parameters coming from Hadamard's
  regularization. However, dimensional regularization will be
  implemented in the present problem not in the general case of an
  arbitrary space-time dimension $D$ but only in the limit where
  $D\rightarrow 4$.
\end{enumerate}
\noindent
Dimensional regularization was invented by 't Hooft and
Veltman~\cite{tHooft, Bollini, Breitenlohner, Collins} as a mean to
preserve the gauge symmetry of perturbative quantum field theories. Our
basic problem here is to respect the gauge symmetry associated with the
diffeomorphism invariance of the classical general relativistic
description of interacting point masses. Hence, we use dimensional
regularization not merely as a trick to compute some particular
integrals which would otherwise be divergent, but as a powerful tool for
solving in a consistent way the Einstein field equations with singular
point-mass sources, while preserving its crucial symmetries. In doing
this, we implicitly assume that the correct theory is the Einstein
general relativity in $D$ space-time dimensions (Section~\ref{secII}).

Earlier work on the equations of motion of point masses at the 2PN
approximation level~\cite{D83houches} was based on the Riesz analytical
continuation method~\cite{Riesz}, which consists of replacing the
delta-function stress-energy tensor of point particles by an auxiliary,
smoother source defined from the Riesz kernel, depending on a complex
number $A$ (in a normal $4$-dimensional space-time). However, it was
mentioned~\cite{D83houches} that the same final result would be obtained
by considering ordinary delta-function sources but in a space-time of
complex dimension $D=4-A$. In other words, the Riesz analytic
continuation method is equivalent to dimensional regularization. It was
also noticed~\cite{D83houches} that the generalization of the Riesz
continuation method to higher post-Newtonian orders is not
straightforward because of the appearance of poles, proportional to
$A^{-1}=(4-D)^{-1}$ at the 3PN order.

In the meantime all calculations were performed using the more
rudimentary Hadamard regularization~\cite{BF00, BFeom, BIJ02, BI04mult},
yielding almost complete results at the 3PN order, \textit{i.e.}
complete but for a few ambiguity parameters, which turn out to be in
fact associated with the latter poles at the 3PN order. Then it was
shown~\cite{DJSdim} how to use dimensional regularization within the ADM
canonical formalism of general relativity at the 3PN order for the
problem of the equations of motion. Further work using dimensional
regularization~\cite{BDE04, BDEI04, BDEI05dr}, finally determined the
values of all the ambiguity parameters, both in the 3PN equations of
motion (parameter called $\lambda$) and in the 3PN gravitational
radiation field (parameters $\xi$, $\kappa$ and $\zeta$). Further in the
same context, we quote Ref.~\cite{GR06} for an alternative approach,
based on a Feynman diagram expansion, showing how to renormalize using
dimensional regularization.

\section{Einstein's Field Equations in $D$ dimensions}\label{secII}
The field equations of general relativity (in $D$-dimensional space-time
with signature is $-+\cdots +$) form a system of ten second-order
partial differential equations obeyed by the space-time metric
$g_{\alpha\beta}$,
\begin{equation}
  E^{\alpha\beta}[g,\partial g,\partial^2g] =
  \frac{8\pi G}{c^4} T^{\alpha\beta}[g],
  \label{EE}
\end{equation}
where the Einstein tensor $E^{\alpha\beta}\equiv
R^{\alpha\beta}-\frac{1}{2}R \, g^{\alpha\beta}$ is generated, through
the gravitational coupling constant $\kappa=8\pi G/c^4$, by the matter
stress-energy tensor $T^{\alpha\beta}$. The gravitational constant $G$
is related to the usual three-dimensional Newton's constant $G_N$ by
\begin{equation}\label{G}
G=G_N\,\ell_0^{d-3},
\end{equation}
where $\ell_0$ denotes an arbitrary length scale. Among the ten Einstein
equations, four govern, \textit{via} the contracted Bianchi identity, the
evolution of the matter system,
\begin{equation}
  \nabla_\mu E^{\alpha\mu}\equiv 0
  \quad \Longrightarrow \quad
  \nabla_\mu T^{\alpha\mu}=0.
  \label{EOM}
\end{equation}
The space-time geometry is constrained by the six remaining equations,
which place six independent constraints on the ten components of the
metric $g_{\alpha\beta}$, leaving four of them to be fixed by a choice
of a coordinate system.

In this paper we adopt the conditions of \textit{harmonic}, or de
Donder, coordinates. We define, as a basic variable, the
gravitational-field amplitude\footnote{Here $g^{\alpha\beta}$ denotes
the contravariant metric (satisfying
$g^{\alpha\mu}g_{\mu\beta}=\delta^\alpha_\beta$), where $g$ is the
determinant of the covariant metric, $g =
\mathrm{det}(g_{\alpha\beta})$, and where $\eta^{\alpha\beta}$
represents an auxiliary Minkowskian metric.}
\begin{equation}
  h^{\alpha\beta} = \sqrt{-g}\, g^{\alpha\beta} - \eta^{\alpha\beta}.
  \label{h}
\end{equation}
The harmonic-coordinate condition, which accounts exactly for the four
equations~(\ref{EOM}) corresponding to the conservation of the matter
tensor, reads
\begin{equation}
  \partial_\mu h^{\alpha\mu} = 0.
  \label{dh}
\end{equation}
The equations~(\ref{h}, \ref{dh}) introduce into the definition of our
coordinate system a preferred Minkowskian structure, with Minkowski
metric $\eta_{\alpha\beta}$. Of course, this is not contrary to the
spirit of general relativity, where there is only one physical metric
$g_{\alpha\beta}$ without any flat prior geometry, because the
coordinates are not governed by geometry (so to speak), but rather are
chosen by researchers when studying physical phenomena and doing
experiments. Actually, the coordinate condition~(\ref{dh}) is especially
useful when we view the gravitational waves as perturbations of
space-time propagating on the fixed Minkowskian manifold with the
background metric $\eta_{\alpha\beta}$. This view is perfectly
legitimate and represents a fruitful and rigorous way to think of the
problem when using approximation methods. Indeed, the metric
$\eta_{\alpha\beta}$, originally introduced in the coordinate
condition~(\ref{dh}), does exist at any \textit{finite} order of
approximation (neglecting higher-order terms), and plays in a sense the
role of some ``prior'' flat geometry.

The Einstein field equations in $D$ dimensions, relaxed by the condition
of harmonic coordinates~(\ref{dh}), can be written in the form of
inhomogeneous flat d'Alembertian equations,
\begin{equation}
  \Box h^{\alpha\beta} = \frac{16\pi G}{c^4} \tau^{\alpha\beta},
  \label{boxh}
\end{equation}
where $\Box\equiv\Box_\eta =\eta^{\mu\nu}\partial_\mu\partial_\nu$
denotes the $D$-dimensional flat space-time d'Alembertian operator. The
source term, $\tau^{\alpha\beta}$, can rightly be interpreted as the
stress-energy pseudo-tensor (actually, $\tau^{\alpha\beta}$ is a Lorentz
tensor) of the matter fields, described by $T^{\alpha\beta}$,
\textit{and} the gravitational field, given by the gravitational source
term $\Lambda^{\alpha\beta}$, \textit{i.e.}
\begin{equation}
  \tau^{\alpha\beta} = |g| T^{\alpha\beta}+ \frac{c^4}{16\pi
  G}\Lambda^{\alpha\beta}. \label{tau}
\end{equation}
The exact expression of $\Lambda^{\alpha\beta}$, including all
non-linearities, reads
\begin{eqnarray}
  \Lambda^{\alpha\beta} = &-& h^{\mu\nu} \partial^2_{\mu\nu}
  h^{\alpha\beta}+\partial_\mu h^{\alpha\nu} \partial_\nu h^{\beta\mu}
  +\frac{1}{2}g^{\alpha\beta}g_{\mu\nu}\partial_\lambda h^{\mu\tau}
  \partial_\tau h^{\nu\lambda} \nonumber \\
  &-&g^{\alpha\mu}g_{\nu\tau}\partial_\lambda h^{\beta\tau} \partial_\mu
  h^{\nu\lambda} -g^{\beta\mu}g_{\nu\tau}\partial_\lambda h^{\alpha\tau}
  \partial_\mu h^{\nu\lambda} +g_{\mu\nu}g^{\lambda\tau}\partial_\lambda
  h^{\alpha\mu} \partial_\tau h^{\beta\nu} \nonumber \\
  &+&\frac{1}{4}(2g^{\alpha\mu}g^{\beta\nu}-g^{\alpha\beta}g^{\mu\nu})
  \Bigl(g_{\lambda\tau}g_{\epsilon\pi}-\frac{1}{D-2}
  g_{\tau\epsilon}g_{\lambda\pi}\Bigr) \partial_\mu h^{\lambda\pi}
  \partial_\nu h^{\tau\epsilon}. \label{Lambda}
\end{eqnarray}
In this form the only explicit dependence on the dimension $D$ is in the
last term of~(\ref{Lambda}). As is clear from this expression,
$\Lambda^{\alpha\beta}$ is made of terms at least quadratic in the
gravitational-field strength $h$ and its first and second space-time
derivatives. In the following, for the highest post-Newtonian order that
we consider (3PN), we need the quadratic, cubic and quartic pieces of
$\Lambda^{\alpha\beta}$. With obvious notation, we can write them as
\begin{equation}
  \Lambda^{\alpha\beta} = N^{\alpha\beta} [h, h] + M^{\alpha\beta} [h,
  h, h] + L^{\alpha\beta}[h, h, h, h] + \mathcal{O}(h^5).
  \label{14_1}
\end{equation}
These various terms can be straightforwardly computed from
Eq.~(\ref{Lambda}); see Eqs.~(3.8) in Ref.~\cite{BFeom} for explicit
expressions.

As said above, the condition~(\ref{dh}) is equivalent to the matter
equations of motion, in the sense of the conservation of the total
pseudo-tensor $\tau^{\alpha\beta}$,
\begin{equation}
  \partial_\mu \tau^{\alpha\mu}=0
  \quad \Longleftrightarrow \quad
  \nabla_\mu T^{\alpha\mu}=0.
  \label{dtau}
\end{equation}
When developing post-Newtonian approximations, we look for the solutions
of the field equations~(\ref{boxh}, \ref{tau}, \ref{Lambda},
\ref{dtau}) under the following four hypotheses:
\begin{enumerate}
\item The matter stress-energy tensor $T^{\alpha\beta}$ is of spatially
  compact support, \textit{i.e.} can be enclosed into some time-like
  world tube, say $r\leq a$, where $r=|\mathbf{x}|$ is the
  harmonic-coordinate radial distance. Outside the domain of the source,
  when $r> a$, the gravitational source term, according to
  Eq.~(\ref{dtau}), is divergence-free,
  \begin{equation}
    \partial_\mu \Lambda^{\alpha\mu} = 0 \qquad (\mathrm{when\ }r>a).
    \label{dLambda}
  \end{equation} 
\item The matter distribution inside the source is smooth:
  $T^{\alpha\beta}\in C^\infty ({\mathbb{R}}^d)$ where $d=D-1$ is the
  space dimension. We have in mind a smooth hydrodynamical ``fluid''
  system, without any singularities nor shocks (\textit{a priori}), that
  is described by some Eulerian equations including high relativistic
  corrections.
\item The source is post-Newtonian in the sense of the existence of the
  small post-Newtonian parameter $v/c=\mathcal{O}(1/c)$. For such a
  source we assume the legitimacy of the method of matched asymptotic
  expansions for matching the inner post-Newtonian field, which is valid
  only in the source's near zone, and the outer multipolar decomposition
  in the source's exterior near zone.
\item The gravitational field has been independent of time (stationary)
  in some remote past, \textit{i.e.} before some finite instant
  $-\mathcal{T}$ in the past, in the sense that
  \begin{equation}
    \frac{\partial}{\partial t}
    \left[h^{\alpha\beta}(\mathbf{x}, t)\right] = 0
    \quad \mathrm{when\ } t\leq -\mathcal{T}.
    \label{past}
  \end{equation}
\end{enumerate}
The latter condition is a mean to impose (somewhat by brute force), the
famous \textit{no-incoming} radiation condition, ensuring that the
matter source is isolated from the rest of the Universe and does not
receive any radiation from infinity. Ideally, the no-incoming radiation
condition should be imposed at past null infinity. One can
argue~\cite{Bliving} that the condition of stationarity in the
past~(\ref{past}), although much weaker than the real no-incoming
radiation condition, does not entail any physical restriction on the
validity of the formulas we derive.

Subject to the condition~(\ref{past}), the Einstein differential field
equations~(\ref{boxh}) can be written equivalently into the form of the
integro-differential equations
\begin{equation}\label{hsol}
h^{\alpha\beta}(\mathbf{x},t) = \frac{16\pi G}{c^4} \int
d^d\mathbf{x}'\,dt'\,G_\mathrm{ret}(\mathbf{x}-\mathbf{x}',t-t')
\,\tau^{\alpha\beta}(\mathbf{x}',t').
\end{equation}
where $G_\mathrm{ret}(\mathbf{x},t)$ is the scalar retarded Green
function in $D=d+1$ dimensions. The Green function for general $d$ has
no simple expression in $(t,\mathbf{x})$ space. However, starting from
its well-known Fourier-space expression, one can write the following
simple integral expression (see \textit{e.g.}~\cite{Cardoso}),
\begin{equation}\label{green}
G_\mathrm{Ret}(\mathbf{x},t)=-\frac{\theta(t)}{(2\pi)^{d/2}}
\int_0^{+\infty}dk \left(\frac{k}{r}\right)^{\frac{d}{ 2}-1}\sin
(c\,k\,t)\,J_{\frac{d}{2}-1}(k\,r).
\end{equation}
Notice that this is in fact a function of $t$ and
$r\equiv\vert\mathbf{x}\vert$ only. Here $\theta(t)$ is the Heaviside
step function, and $J_{\frac{d}{2}-1}(k\,r)$ the usual Bessel function
[see Eq.~(\ref{Bessel}) in the Appendix].

\section{Dimensional regularization of the equations of motion}\label{secIII}
As said before, work at the 3PN order using Hadamard's self-field
regularization showed the appearance of ambiguity parameters, due to an
incompleteness of the Hadamard regularization employed for curing the
infinite self-field of point particles. By ambiguity parameter, we mean
a dimensionless coefficient which cannot be computed within the Hadamard
regularization scheme. Nevertheless, the majority of terms could be
computed unambiguously with Hadamard's regularization~\cite{BF00,BFeom}.
We summarize here the determination using dimensional regularization of
the ambiguity parameter $\lambda$ which appeared in the 3PN equations of
motion; note that $\lambda$ is equivalent to the static ambiguity
parameter $\omega_\mathrm{static}$ originally introduced in
Refs.~\cite{JaraS98, JaraS99}.

The post-Newtonian iteration of the Einstein field equations with
point-like matter source (Dirac delta-functions with spatial supports
$\mathbf{y}_1$ and $\mathbf{y}_2$) yields a generic form for the
functions representing the metric components in successive
post-Newtonian approximations. The generic functions we have to deal
with in $3$ dimensions, say $F(\mathbf{x})$, are smooth on
$\mathbb{R}^3$ except at $\mathbf{y}_1$ and $\mathbf{y}_2$, around which
they admit singular Laurent-type expansions in powers and inverse powers
of $r_1\equiv\vert\mathbf{x}-\mathbf{y}_1\vert$ and
$r_2\equiv\vert\mathbf{x}-\mathbf{y}_2\vert$, given by (say, for any
$N$)\footnote{The function $F(\mathbf{x})$ depends also on time $t$,
through for instance its dependence on the velocities $\mathbf{v}_1(t)$
and $\mathbf{v}_2(t)$, but the (coordinate) $t$ time is purely
``spectator'' in the regularization process, and thus will not be
indicated.}
\begin{equation}\label{Fx}
F(\mathbf{x})=\sum_{p_0\leq p\leq N}r_1^{p}
\mathop{f}_1{}_{p}(\mathbf{n} _1)+o(r_1^N),
\end{equation}
and similarly for the other point 2. Here
$r_1=|\mathbf{x}-\mathbf{y}_1|\rightarrow 0$, and the coefficients
${}_1f_p$ of the various powers of $r_1$ depend on the unit direction
$\mathbf{n}_1=(\mathbf{x}-\mathbf{y}_1)/r_1$ of approach to the singular
point.\footnote{The $o$ Landau symbol for remainders takes its standard
meaning.} The powers $p$ of $r_1$ are relative integers, and are bounded
from below ($p_0\leq p$). The coefficients ${}_1f_p$ (and ${}_2f_p$) for
which $p<0$ can be referred to as the \textit{singular} coefficients of
$F$.

A function $F$ defined in $3$ dimensions being given, and admitting the
singular expansion~(\ref{Fx}), we define the Hadamard
\textit{partie finie} of $F$ at the location of the particle 1, where it is
singular, as
\begin{equation}
  (F)_1= \int \frac{d\Omega_1}{ 4\pi}\,\mathop{f}_{1}{}_{\!\!0}(\mathbf{n}_1),
  \label{F1}
\end{equation}
where $d\Omega_1$ denotes the solid angle element centered on
$\mathbf{y}_1$ and of direction $\mathbf{n}_1$. An important feature we
have to notice is that because of the angular integration in
Eq.~(\ref{F1}), the Hadamard partie finie is ``non-distributive'' in the
sense that $(FG)_1\not= (F)_1(G)_1$ in general. The non-distributivity
of Hadamard's partie finie will be the main source of the appearance of
ambiguity parameters at the 3PN order; remarkably it does not affect any
calculation before that order. The second notion of Hadamard partie
finie (in short $\mathrm{Pf}$) concerns that of the integral $\int
d^3\mathbf{x} \, F$, which is generically divergent at the location of
the two singular points $\mathbf{y}_1$ and $\mathbf{y}_2$ (we assume for
the moment that the integral converges at infinity). It is defined by
\begin{equation}
  \mathrm{Pf}_{s_1 s_2} \int d^3\mathbf{x} \, F = \lim_{s \rightarrow 0}
  \, \biggl\{\int_{\mathcal{S}(s)} d^3\mathbf{x} \, F + 4\pi\sum_{a+3<
  0}{\frac{s^{a+3}}{a+3}} \left( \frac{F}{r_1^a} \right)_1 + 4 \pi \ln
  \left(\frac{s}{s_1}\right) \left(r_1^3 F\right)_1 + 1\leftrightarrow
  2\biggr\}. \label{PfF}
\end{equation}
The first term integrates over a domain $\mathcal{S}(s)$ defined as
${\mathbb{R}}^3$ from which the two spherical balls $r_1\leq s$ and
$r_2\leq s$ of radius $s$ and centered on the two singularities are
excised. The other terms, where the value of a function at point 1 takes
the meaning~(\ref{F1}), are such that they cancel out the divergent part
of the first term in the limit where $s\rightarrow 0$ (the symbol
$1\leftrightarrow 2$ means the same terms but corresponding to the other
point 2). The Hadamard partie-finie integral depends on two strictly
positive constants $s_1$ and $s_2$, associated with the logarithms
present in Eq.~(\ref{PfF}). 

The post-Newtonian approximation consists of breaking the hyperbolic
d'Alembertian operator $\Box$ in Eq.~(\ref{boxh}) into the elliptic
Laplacian operator $\Delta$ plus the famous retardation term
$-c^{-2}\partial_t^2$ which is to be considered small in the
post-Newtonian sense, and put in the RHS of the equation where it is
iterated. As a consequence, we essentially have to deal with the
regularization of Poisson integrals, or iterated Poisson integrals, of
the generic function $F$. In the case of a Poisson integral potential in
$3$ dimensions, say
\begin{equation}
P({\mathbf{x}}')= -\frac{1}{4\pi}
\int\frac{d^3{\mathbf{x}}}{\vert{\mathbf{x}}-{\mathbf{x}}'\vert}
F({\mathbf{x}}),
\label{Px}
\end{equation}
the Hadamard partie finie integral must be defined in a more precise
way. Indeed, the definition~(\ref{F1}) \textit{stricto sensu} is
applicable when the expansion of the function $F$, when $r_1\rightarrow
0$, does not involve logarithms of $r_1$; see Eq.~(\ref{Fx}). However,
the Poisson integral $P(\mathbf{x}')$ of $F(\mathbf{x})$ will typically
involve such logarithms (these will appear precisely at the 3PN order),
namely some $\ln r'_1$ where $r'_1\equiv
\vert\mathbf{x}'-\mathbf{y}_1\vert$ tends to zero (hence $\ln
r'_1$ is formally infinite). The proper way to define the Hadamard
partie finie in this case is to include the $\ln r'_1$ into its
definition, and we arrive at~\cite{BFreg}
\begin{equation}
(P)_1 = -\frac{1}{4\pi}{\mathrm{Pf}}_{r_1',s_2}
\int\frac{d^3{\mathbf{x}}}{r_1} F({\mathbf{x}}) - (r_1^2\,F)_1.
\label{P1}
\end{equation}
The first term follows from Hadamard's partie finie
integral~(\ref{PfF}); the second one is given by Eq.~(\ref{F1}). Notice
that in this result the constant $s_1$ entering the partie finie
integral~(\ref{PfF}) has been ``replaced'' by $r'_1$, which plays the
role of a new regularization constant (together with $r'_2$ for the
other particle), and which ultimately parametrizes the final Hadamard
regularized 3PN equations of motion. It was shown that $r'_1$ and $r'_2$
are unphysical, in the sense that they can be removed by a coordinate
transformation~\cite{BF00, BFeom}. On the other hand, the constant $s_2$
remaining in the result~(\ref{P1}) is the source for the appearance of
the physical ambiguity parameter, called $\lambda$, as it will be
related to it by Eq.~(\ref{lnr2s2}) below.

In $d$ spatial dimensions, there is an analogue of the function $F$,
which results from the same post-Newtonian iteration process but
performed in $d$ spatial dimensions. Let us call this function
$F^{(d)}(\mathbf{x})$, where $\mathbf{x}\in\mathbb{R}^d$. When
$r_1\rightarrow 0$ the function $F^{(d)}$ admits a singular expansion
which is richer than in $3$ dimensions. Posing $\varepsilon = d-3$ we
have
\begin{equation}\label{Fdx}
F^{(d)}(\mathbf{x})=\sum_{\substack{p_0\leq p\leq N\\ q_0\leq q\leq
q_1}}r_1^{p+q\varepsilon}
\mathop{f}_1{}_{p,q}^{(\varepsilon)}(\mathbf{n} _1)+o(r_1^N).
\end{equation}
The coefficients $\mathop{f}_1{}_{p,q}^{(\varepsilon)}(\mathbf{n}_1)$
depend on the dimension; the powers of $r_1$ involve the relative
integers $p$ and $q$ whose values are limited by some $p_0$, $q_0$ and
$q_1$ as indicated. The Poisson integral of $F^{(d)}$, in $d$
dimensions, is given by the Green's function for the Laplace operator,
\begin{equation}
P^{(d)}({\mathbf{x}}')= -\frac{\tilde{k}}{4\pi}
\int\frac{d^d{\mathbf{x}}}{\vert{\mathbf{x}}-{\mathbf{x}}'\vert^{d-2}}
F^{(d)}({\mathbf{x}}),
\label{Pdx}
\end{equation}
where $\tilde{k}$ is a constant related to the usual Eulerian
$\Gamma$-function by
\begin{equation}\label{ktilde}
\tilde{k}=\frac{\Gamma\left(\frac{d-2}{2}\right)}{\pi^{
\frac{d-2}{2}}}.
\end{equation}
We have $\lim_{d\rightarrow 3}\tilde{k}=1$. Notice also that $\tilde{k}$
is closely linked to the volume $\Omega_{d-1}$ of the sphere with $d-1$
dimensions; see Eq.~(\ref{ktildeBis}) in the Appendix.

We need to evaluate the Poisson integral at the point ${\mathbf{x}}' =
{\mathbf{y}}_1$ where it is singular; in contrast with the case of
Hadamard regularization where the result is given by~(\ref{P1}), this is
quite easy in dimensional regularization, because the nice properties of
analytic continuation allow simply to get $[P^{(d)}
({\mathbf{x}}')]_{\mathbf{x}' = {\mathbf{y}}_1}$ by replacing
${\mathbf{x}}'$ by ${\mathbf{y}}_1$ into the explicit integral form
(\ref{Pdx}). So we simply have
\begin{equation}
P^{(d)}({\mathbf{y}}_1)=-\frac{\tilde{k}}{4\pi}
\int\frac{d^d{\mathbf{x}}}{r_1^{d-2}}F^{(d)}({\mathbf{x}}).\label{Pd}
\end{equation}
The main technical step of our strategy to compute the ambiguity
parameter $\lambda$ will consist of computing, in the limit
$\varepsilon\rightarrow 0$, the \textit{difference} between the
$d$-dimensional Poisson potential~(\ref{Pd}), and its $3$-dimensional
counterpart which is defined from Hadamard's self-field regularization
and given by~(\ref{P1}). Denoting the difference between the dimensional
and Hadamard regularizations by means of the script letter
$\mathcal{D}$, we pose (for the result concerning the point 1)
\begin{equation}
\mathcal{D}P_1\equiv P^{(d)}({\mathbf{y}}_1)-(P)_1.
\label{DP1}
\end{equation}
That is, $\mathcal{D}P_1$ is what we shall have to \textit{add} to the
Hadamard-regularization result in order to get the $d$-dimensional
result. However, we shall only compute the first two terms of the
Laurent expansion of $\mathcal{D}P_1$ when $\varepsilon=d-3 \rightarrow
0$, say $a_{-1} \, \varepsilon^{-1} + a_0 + \mathcal{O} (\varepsilon)$.
This is the information we need to determine the value of the ambiguity
parameter. Notice that the difference $\mathcal{D}P_1$ comes exclusively
from the contribution of terms developing some poles $\propto
1/\varepsilon$ in the $d$-dimensional calculation.

Let us next outline the way we obtain, starting from the computation of
the ``difference'', the 3PN equations of motion in dimensional
regularization, and show how the ambiguity parameter $\lambda$ is fixed
by the process. By contrast to $r'_1$ and $r'_2$ which are pure gauge,
$\lambda$ is a genuine physical ambiguity, introduced in
Refs.~\cite{BFreg, BFeom} as the \textit{single} unknown numerical
constant parametrizing the ratio between $s_2$ and $r'_2$ [where $s_2$
is the constant left in Eq.~(\ref{P1})] as
\begin{equation}
\ln\Bigl(\frac{r_2'}{s_2}\Bigr)=\frac{159}{308}+\lambda
\frac{m_1+m_2}{m_2}\quad\text{(and $1\leftrightarrow 2$)},
\label{lnr2s2}
\end{equation}
where $m_1$ and $m_2$ are the two masses. The terms corresponding to the
$\lambda$-ambiguity in the acceleration $\mathbf{a}_1=d\mathbf{v}_1/dt$
of particle 1 read simply 
\begin{equation}
\Delta\mathbf{a}_1 [\lambda] =
-\frac{44\lambda}{3}\,\frac{G_N^4\,m_1\,m_2^2\,(m_1+m_2)}{r_{12}^5\,c^6}
\,\mathbf{n}_{12},
\label{Dealtaa1}
\end{equation}
where the relative distance between particles is denoted
$\mathbf{y}_1-\mathbf{y}_2\equiv r_{12}\,\mathbf{n}_{12}$ (with
$\mathbf{n}_{12}$ the unit vector pointing from particle 2 to particle
1). We start from the end result of Ref.~\cite{BFeom} for the 3PN
harmonic coordinates acceleration $\mathbf{a}_1$ in Hadamard's
regularization, abbreviated as HR. Since the result was in fact obtained
by means of a specific variant of HR, called the extended Hadamard's
regularization (in short EHR), we write it as
\begin{equation}
\mathbf{a}_1^\mathrm{(HR)} = \mathbf{a}_1^\mathrm{(EHR)} + \Delta\mathbf{a}_1
[\lambda],
\label{a1HR}
\end{equation}
where $\mathbf{a}_1^\mathrm{(EHR)}$ is a fully determined functional of
the masses $m_1$ and $m_2$, the relative distance
$r_{12}\,\mathbf{n}_{12}$, the coordinate velocities $\mathbf{v}_1$ and
$\mathbf{v}_2$, and also the gauge constants $r_1'$ and $r_2'$. The only
ambiguous term is the second one and is given by Eq.~(\ref{Dealtaa1}).

Our method is to express both the dimensional and Hadamard
regularizations in terms of their common ``core'' part, obtained by
applying the so-called ``pure-Hadamard-Schwartz'' (pHS) regularization.
Following the definition of Ref.~\cite{BDE04}, the pHS regularization is
a specific, minimal Hadamard-type regularization of integrals, based on
the partie finie integral~(\ref{PfF}), together with a minimal treatment
of ``contact'' terms, in which the definition~(\ref{PfF}) is applied
separately to each of the elementary potentials, denoted $V$, $V_i$,
$\hat{W}_{ij}$ $\cdots$, that enter the post-Newtonian metric. The pHS
regularization also assumes the use of standard Schwartz distributional
derivatives~\cite{Schwartz}. The interest of the pHS regularization is
that the dimensional regularization is equal to it plus the
``difference''; see Eq.~(\ref{a1DimReg}) below.

To obtain the pHS-regularized acceleration we need to substract from the
EHR result a series of contributions, which are specific consequences of
the use of EHR~\cite{BFreg,BFregM}. For instance, one of these
contributions corresponds to the fact that in the EHR the distributional
derivative differs from the usual Schwartz distributional derivative.
Hence we define
\begin{equation}
{\mathbf{a}}_1^\mathrm{(pHS)} = {\mathbf{a}}_1^\mathrm{(EHR)} -
\sum_{A}\delta_A{\mathbf{a}}_1,
\label{accpH}
\end{equation}
where the $\delta_A{\mathbf{a}}_1$'s denote the extra terms following
from the EHR prescriptions. The pHS-regularized
acceleration~(\ref{accpH}) constitutes essentially the result of the
first stage of the calculation of ${\mathbf{a}}_1$, composed of plenty
of terms and which are all perfectly well-defined.

The next step consists of evaluating the Laurent expansion, in powers of
$\varepsilon = d-3$, of the difference between the dimensional
regularization and the pHS ($3$-dimensional) computation. As we said
above this difference makes a contribution only when a term generates a
\textit{pole} $\sim 1/\varepsilon$, in which case the dimensional
regularization adds an extra contribution, made of the pole and the
finite part associated with the pole [we consistently neglect all terms
$\mathcal{O}(\varepsilon)$]. One must then be especially wary of
combinations of terms whose pole parts finally cancel (``cancelled
poles'') but whose dimensionally regularized finite parts generally do
not, and must be evaluated with care. We denote the above defined
difference by
\begin{equation}
\mathcal{D}{\mathbf{a}}_1 = \sum \mathcal{D}P_1.
\label{deltaacc}
\end{equation}
It is made of the sum of all the individual differences of Poisson or
Poisson-like integrals as computed in Eq.~(\ref{DP1}). The total
difference~(\ref{deltaacc}) depends on the Hadamard regularization
scales $r_1'$ and $s_2$ (or equivalently on $\lambda$ and $r_1'$,
$r_2'$), and on the parameters associated with dimensional
regularization, namely $\varepsilon$ and the characteristic length scale
$\ell_0$ introduced in Eq.~(\ref{G}). Finally, our main result is the
explicit computation of the dimensional regularization (DR) acceleration
as
\begin{equation}
{\mathbf{a}}_1^\mathrm{(DR)} = {\mathbf{a}}_1^\mathrm{(pHS)} + {\cal
D}{\mathbf{a}}_1.
\label{a1DimReg}
\end{equation}
With this result we can prove two theorems~\cite{BDE04}.

\begin{theorem}
The pole part $\propto 1/\varepsilon$ of the DR
acceleration~(\ref{a1DimReg}) can be re-absorbed (\textit{i.e.},
renormalized) into some shifts of the two ``bare'' world-lines:
$\mathbf{y}_1 \rightarrow \mathbf{y}_1+\bm{\xi}_1$ and $\mathbf{y}_2
\rightarrow \mathbf{y}_2+\bm{\xi}_2$, with, say, $\bm{\xi}_{1,2} \propto
1/\varepsilon$, so that the result, expressed in terms of the
``dressed'' quantities, is finite when $\varepsilon\rightarrow 0$.
\label{th1}
\end{theorem}

\noindent
The precise shifts $\bm{\xi}_1$ and $\bm{\xi}_2$ involve not only a pole
contribution $\propto 1/\varepsilon$ [which would define a
renormalization by minimal subtraction (MS)], but also a finite
contribution when $\varepsilon\rightarrow 0$. Their explicit expressions
read:
\begin{equation}
\bm{\xi}_1=\frac{11}{3}\frac{G_N^2\,m_1^2}{c^6}\left[
\frac{1}{\varepsilon}-2\ln\left(
\frac{r'_1\overline{q}^{1/2}}{\ell_0}\right) -\frac{327}{1540}\right]
{\mathbf{a}}_{N1}~~\text{(together with $1\leftrightarrow 2$)},
\end{equation}
where $G_N$ is Newton's constant, $\ell_0$ is the characteristic length
scale of dimensional regularization [\textit{cf.} Eq.~(\ref{G})],
${\mathbf{a}}_{N1}$ is the Newtonian acceleration of the particle 1 in
$d$ dimensions, and $\overline{q}\equiv 4\pi e^C$ depends on Euler's
constant $C=0.577\cdots$. Note that when working at the level of the
equations of motion (not considering the metric outside the
world-lines), the effect of shifts can be seen as being induced by a
coordinate transformation of the bulk metric as in Ref.~\cite{BFeom}.

\begin{theorem}
The renormalized (finite) DR acceleration is physically equivalent to
the Hadamard-regularized (HR) acceleration (end result of
Ref.~\cite{BFeom}), in the sense that
\begin{equation}
{\mathbf{a}}_1^{\mathrm{(HR)}} = \lim_{\varepsilon\rightarrow 0} \,
\bigl[{\mathbf{a}}_1^{\mathrm{(DR)}} + \delta_{\bm{\xi}} \,
{\mathbf{a}}_1 \bigr],
\label{eta}
\end{equation}
where $\delta_{\bm{\xi}} \, {\mathbf{a}}_1$ denotes the effect of the
shifts on the acceleration, if and only if the HR ambiguity parameter
$\lambda$ entering the harmonic-coordinates equations of motion takes
the unique value
\begin{equation}\label{lambda}
\lambda = -\frac{1987}{3080}.
\end{equation}
\label{th2}
\end{theorem}

\noindent
The coefficient $\lambda$ has therefore been fixed by dimensional
regularization, both within the ADM-Hamiltonian formalism~\cite{DJSdim},
and the harmonic-coordinates equations of motion~\cite{BDE04}.

An alternative work, by Itoh and Futamase~\cite{itoh1, itoh2}, following
previous investigations in Refs.~\cite{IFA00, IFA01}, has derived the
3PN equations of motion in harmonic coordinates by means of a variant of
the famous ``surface-integral'' method introduced by Einstein, Infeld
and Hoffmann~\cite{EIH}. The aim is to describe extended relativistic
compact binary systems in the strong-field point particle
limit~\cite{F87}. This approach, alternative to the use of self-field
regularizations, is interesting because it is based on the physical
notion of extended compact bodies in general relativity, and is free of
the problems of ambiguities. The end result of Refs.~\cite{itoh1, itoh2}
is in agreement with the 3PN harmonic coordinates equations of
motion~\cite{BF00, BFeom} and, moreover, it does determine the ambiguity
parameter $\lambda$ to exactly the value~(\ref{lambda}).

\section{Dimensional regularization of the radiation field}\label{secIV}
We now address the similar problem concerning the binary's gravitational
radiation field (3PN beyond the Einstein quadrupole formalism), for
which three ambiguity parameters, denoted $\xi$, $\kappa$ and $\zeta$,
have been shown to appear due to the Hadamard self-field
regularization~\cite{BIJ02, BI04mult}. To apply dimensional
regularization, we must use as we did for the equations of motion in
Section~\ref{secIII} the $d$-dimensional post-Newtonian iteration; and,
crucially, we have to generalize to $d$ dimensions some key results from
the gravitational wave generation formalism. The specific wave
generation formalism we employ is based on a post-Newtonian expansion
for the metric field in the near zone of the source, and on the
so-called multipolar-post-Minkowskian expansion for the field in the
exterior of the source, including the regions at infinity from the
source where the detector is located~\cite{BD86}. The expression of the
multipole moments describing the physical (post-Newtonian) source are
then obtained by a technique of asymptotic matching performed in the
overlapping region of common validity between the two types of
expansion, namely the exterior part of the near zone~\cite{B95,B98mult}
(see~\cite{Bliving} for a review).

Let us first recall the expressions of the source multipole moments
$I_L$ (mass-type moment) and $J_L$ (current-type) of an isolated
post-Newtonian source in ordinary $3$-dimensional space. They are given,
for multipolarities $\ell\geq 2$, by~\cite{B98mult}\footnote{Our
notation is the following: $L=i_1i_2\dots i_l$ denotes a multi-index,
made of $l$ spatial indices; similarly we write for instance
$aL=ai_1\dots i_{l}$ or $L-1=i_1\dots i_{l-1}$. The symmetric-trace-free
(STF) projection is denoted with a hat, so that ${\hat x}_L=x_{\langle
L\rangle}$ is the STF projection of the product of $l$ spatial vectors,
denoted $x_L=x_{i_1}\dots x_{i_l}$. Sometimes we also indicate the STF
projection by brackets surrounding indices ${\hat x}_L=x_{\langle
L\rangle}$. Note that an expansion into STF tensors ${\hat n}_L={\hat
x}_L/r^l$ (which are functions of the spherical angles $\theta$ and
$\phi$) is equivalent to the usual expansion in spherical harmonics
$\mathrm{Y}_{lm}(\theta,\phi)$. The dots indicate successive
time-derivations.}
\begin{subequations}\label{ILJL}\begin{eqnarray}
I_L(t)&=& \mathcal{F\!\!P}\,\int d^3\mathbf{x}\,\int^1_{-1} dz\left\{
\delta_\ell(z)\,\hat{x}_L\,\Sigma
-\frac{4(2\ell+1)}{c^2(\ell+1)(2\ell+3)} \,\delta_{\ell+1}(z)
\,\hat{x}_{iL} \,\dot{\Sigma}_i\right.\nonumber\\ &&\qquad\quad \left.
+\frac{2(2\ell+1)}{c^4(\ell+1)(\ell+2)(2\ell+5)}
\,\delta_{\ell+2}(z)\,\hat{x}_{ijL}\ddot{\Sigma}_{ij}\right\}
(\mathbf{x},t+z \vert{\mathbf{x}}\vert/c),\label{IL}\\J_L(t)&=&
\mathcal{F\!\!P}\,\varepsilon_{ab<i_\ell} \int d^3
\mathbf{x}\,\int^1_{-1} dz\biggl\{ \delta_\ell(z)\,\hat{x}_{L-1>a}
\,\Sigma_b \nonumber\\ &&\qquad\quad
-\frac{2\ell+1}{c^2(\ell+2)(2\ell+3)}
\,\delta_{\ell+1}(z)\,\hat{x}_{L-1>ac} \,\dot{\Sigma}_{bc}\biggr\}
(\mathbf{x},t+z \vert\mathbf{x}\vert/c),\label{JL}
\end{eqnarray}\end{subequations}
where the source densities $\Sigma_{\mu\nu}$'s are evaluated at the
position $\mathbf{x}$ and at time $t+z \vert{\mathbf{x}}\vert/c$, and
are defined by
\begin{equation}
\Sigma = \frac{\overline\tau^{00}+\overline\tau^{ii}}{c^2},\qquad
\Sigma_i = \frac{\overline\tau^{0i}}{c},\qquad \Sigma_{ij} =
\displaystyle \overline{\tau}^{ij}. \label{Sig}
\end{equation}
Here $\tau^{\alpha\beta}$ is the pseudo stress-energy tensor~(\ref{tau})
in $3$ dimensions and the overbar refers to its formal post-Newtonian
expansion,
$\overline\tau^{\alpha\beta}\equiv\mathrm{PN}[\tau^{\alpha\beta}]$. Let
us note that the expressions~(\ref{ILJL}) are ``\textit{exact}'', in the
sense that they are formally valid up to any PN order.
Equations~(\ref{ILJL}) involve an integration over the variable $z$,
with associated function $\delta_\ell (z)$ given by
\begin{equation}\label{deltal}
\delta_\ell (z) \equiv \frac{(2\ell+1)!!}{2^{\ell+1} \ell!}
\,(1-z^2)^\ell, \qquad \int^1_{-1} dz\,\delta_\ell (z) = 1, \qquad
\lim_{\ell \rightarrow +\infty}\delta_\ell (z) = \delta (z),
\end{equation}
where $\delta(z)$ is the usual Dirac's one-dimensional delta-function.
In practice, the post-Newtonian moments~(\ref{ILJL}) are to be computed
by means of the infinite post-Newtonian series
\begin{equation}\label{intdeltal}
\int^1_{-1} dz~ \delta_\ell(z) \,\Sigma(\mathbf{x},t+z
\vert\mathbf{x}\vert/c) =
\sum_{k=0}^{+\infty}\,\frac{(2\ell+1)!!}{(2k)!!(2\ell+2k+1)!!}
\,\left(\frac{\vert\mathbf{x}\vert}{c}\frac{\partial}{\partial
t}\right)^{2k} \!\Sigma(\mathbf{x},t).
\end{equation}

In Eqs.~(\ref{ILJL}) there is a special process of taking the ``finite
part'', indicated by the symbol $\mathcal{F\!\!P}$, which is necessary
in order to deal with the bound of the integral at infinity,
corresponding to infra-red divergencies, in the limit
$\vert\mathbf{x}\vert\rightarrow\infty$. Indeed, notice that the pseudo
stress-energy tensor $\tau^{\alpha\beta}$ includes the crucial
contribution of the gravitational field, denoted $\Lambda^{\alpha\beta}$
in Eq.~(\ref{tau}), which has a spatially \textit{non-compact} support.
This fact, together with the presence of the multipolar factor
$\hat{x}_L$ in the integrand, prevents one to (naively) write the
standard expressions for the multipole moments valid for
compact-supported sources; such expressions have no meaning in
non-linear general relativity. The solution to this dilemna has been to
introduce~\cite{B95, B98mult} the specific finite part
$\mathcal{F\!\!P}$, and to show how these specific multipole moments so
defined are related to the physical wave form at infinity.

To proceed with dimensional regularization, we need the $d$-dimensional
analogues of the multipole moments of the source, say
$\mathrm{I}^{(d)}_L$ and $\mathrm{J}^{(d)}_L$, consequences of the
$D$-dimensional Einstein field equations for isolated post-Newtonian
sources. In the case of the mass-type moments we find~\cite{BDEI05dr}
\begin{eqnarray}
\mathrm{I}^{(d)}_L(t)&=& \frac{d-1}{2(d-2)}\,\mathcal{F\!\!P} \int d^d\mathbf{x}\,
\biggl\{\hat{x}_L\,\mathop{\Sigma}_{[l]}(\mathbf{x},t) -
\frac{4(d+2l-2)}{c^2(d+l-2)(d+2l)} \,\hat
x_{aL}\,\mathop{\dot\Sigma}_{[l+1]}{}_{\!\!a}(\mathbf{x},t)\nonumber\\
&&\qquad+\frac{2(d+2l-2)}{c^4(d+l-1)(d+l-2)(d+2l+2)}\hat{x}_{abL}
\,\mathop{\ddot\Sigma}_{[l+2]}{}_{\!\!ab}(\mathbf{x},t)\biggr\},
\label{ILd}
\end{eqnarray}
where we denote [generalizing Eq.~(\ref{Sig})]
\begin{equation}
\Sigma =
\frac{2}{d-1}\frac{(d-2)\overline\tau^{00}+\overline\tau^{ii}}{c^2},\qquad
\Sigma_i = \frac{\overline\tau^{0i}}{c},\qquad \Sigma_{ij} =
\displaystyle \overline{\tau}^{ij}, \label{Sigd}
\end{equation}
where now $\overline\tau^{\alpha\beta}$ is the post-Newtonian expansion
of the pseudo stress-energy tensor in $D$ dimensions. For any of the
latter source densities the underscript $[l]$ means the infinite series
\begin{eqnarray}\label{series}
\mathop{\Sigma}_{[l]}(\mathbf{x},t) =
\sum_{k=0}^{+\infty}\frac{1}{2^{2k}k!}\frac{\Gamma\left(\frac{d}{2}+
l\right)}{\Gamma\left(\frac{d}{2}+l+k\right)}\left(\frac{\vert
\mathbf{x}\vert}{c}\frac{\partial}{\partial
t}\right)^{2k}\!\!\Sigma(\mathbf{x},t),
\end{eqnarray}
which constitutes the $d$-dimensional version of Eq.~(\ref{intdeltal}).
At Newtonian order Eq.~(\ref{ILd}) reduces to the standard result ${\rm
I}_L^{(d)}=\int d^d\mathbf{x}\,\rho\,\hat{x}_L+\mathcal{O}(c^{-2})$ with
$\rho=T^{00}/c^2$.

Like for the case of the equations of motion, the ambiguity parameters
$\xi$, $\kappa$ and $\zeta$ come from a deficiency of the Hadamard
regularization coming up at 3PN order and mainly due to its
``non-distributivity''. The terms corresponding to these ambiguities are
contained in the 3PN mass quadrupole moment $\mathrm{I}_{ij}$ and are
found to be
\begin{equation}
\Delta \mathrm{I}_{ij}[\xi, \kappa, \zeta] =
\frac{44}{3}\frac{G_N^2\,m_1^3}{c^6}\biggl[\left(\xi + \kappa
\frac{m_1+m_2}{m_1}\right) y_1^{\langle i}a_1^{j\rangle} +\,\zeta
\,v_1^{\langle i}v_1^{j\rangle}\biggr] + 1\leftrightarrow 2,
\label{amb}
\end{equation}
where $\mathbf{y}_1$, $\mathbf{v}_1$ and $\mathbf{a}_1$ denote the first
particle's position, velocity and acceleration (the brackets
$\langle\rangle$ surrounding indices refer to the STF projection). As in
Section~\ref{secIII}, we express the Hadamard and dimensional results in
terms of the more basic pure-Hadamard-Schwartz (pHS) regularization. The
first step of the calculation~\cite{BI04mult} is therefore to relate the
Hadamard-regularized quadrupole moment $I_{ij}^{(\mathrm{HR})}$, for
general orbits, to its pHS part; we find:
\begin{equation}
\mathrm{I}_{ij}^{(\mathrm{HR})} = \mathrm{I}_{ij}^{(\mathrm{pHS})} + \Delta {\rm
I}_{ij}\Bigl[\xi+\frac{1}{22},\kappa,\zeta+\frac{9}{110}\Bigr].
\label{IijH}
\end{equation}
In the RHS we see both the pHS part, and the effect of adding the
ambiguities, with some numerical shifts of the ambiguity parameters
coming from the difference between the specific Hadamard-type
regularization scheme used in Ref.~\cite{BIJ02} and the pHS one. The pHS
part is free of ambiguities but depends on the gauge constants $r_1'$
and $r_2'$ introduced in the harmonic-coordinates equations of
motion~\cite{BF00, BFeom}.

We next use the $d$-dimensional moment~(\ref{ILd}) to compute the
difference between the dimensional regularization (DR) result and the
pHS one~\cite{BDEI04, BDEI05dr}. As in the work on equations of motion,
we find that the ambiguities arise solely from the terms in the
integration regions near the particles (\textit{i.e.},
$r_1=\vert\mathbf{x}-\mathbf{y}_1\vert \rightarrow 0$ or
$r_2=\vert\mathbf{x}-\mathbf{y}_2\vert \rightarrow 0$) that give rise to
poles $\propto 1/\varepsilon$, corresponding to logarithmic ultra-violet
divergences in 3 dimensions. The infra-red region at infinity
(\textit{i.e.}, $\vert\mathrm{x}\vert\rightarrow +\infty$) does not
contribute to the difference DR $-$ pHS. The compact-support terms in
the integrand of~(\ref{ILd}), proportional to the components of the
matter stress-energy tensor $T^{\alpha\beta}$, are also found not to
contribute to the difference. We are therefore left with evaluating the
difference linked with the computation of the \textit{non-compact} terms
in the expansion of the integrand in~(\ref{ILd}) near the singularities
that produce poles in $d$ dimensions.

Let $F^{(d)}(\mathbf{x})$ be the non-compact part of the integrand of
the quadrupole moment~(\ref{ILd}) (with indices $L=ij$), where $F^{(d)}$
includes the appropriate multipolar factors such as $\hat{x}_{ij}$, so
that
\begin{equation}
\mathrm{I}^{(d)}_{ij} = \int d^d\mathbf{x}\,F^{(d)}(\mathbf{x}).
\label{ILFd}
\end{equation}
We do not indicate that we are considering here only the non-compact
part of the moments. Near the singularities the function
$F^{(d)}(\mathbf{x})$ admits a singular expansion of the
type~(\ref{Fdx}). In practice, the coefficients ${}_1f_{p,q}^{
(\varepsilon)}$ are computed by performing explicitly the post-Newtonian
iteration. On the other hand, the analogue of Eq.~(\ref{ILFd}) in $3$
dimensions is
\begin{equation}
\mathrm{I}_{ij} = \mathrm{Pf} \int d^3\mathbf{x}\,F(\mathbf{x}),
\label{ILF}
\end{equation}
where $\mathrm{Pf}$ refers to the Hadamard partie finie defined by
Eq.~(\ref{PfF}). The difference $\mathcal{D}\mathrm{I}$ between the DR
evaluation of the $d$-dimensional integral~(\ref{ILFd}), and its
corresponding three-dimensional evaluation, \textit{i.e.} the partie
finie~(\ref{ILF}), reads then
\begin{equation}
\mathcal{D}\mathrm{I}_{ij} = \mathrm{I}^{(d)}_{ij} - \mathrm{I}_{ij}.
\label{DIL}
\end{equation}
Such difference depends only on the ultra-violet behavior of the
integrands, and can therefore be computed ``locally'', \textit{i.e.} in
the vicinity of the particles, when $r_1 \rightarrow 0$ and $r_2
\rightarrow 0$. We find that Eq.~(\ref{DIL}) depends on two constant
scales $s_1$ and $s_2$ coming from Hadamard's partie finie~(\ref{PfF}),
and on the constants belonging to dimensional regularization, which are
$\varepsilon=d-3$ and the length scale $\ell_0$ defined by
Eq.~(\ref{G}). The dimensional regularization of the 3PN quadrupole
moment is then obtained as the sum of the pHS part, and of the
difference computed according to Eq.~(\ref{DIL}), namely
\begin{equation}\label{IijDR}
\mathrm{I}_{ij}^{(\mathrm{DR})} = \mathrm{I}_{ij}^{(\mathrm{pHS})} +
\mathcal{D}\mathrm{I}_{ij}.
\end{equation}
An important fact, hidden in our too-compact notation~(\ref{IijDR}), is
that the RHS of~(\ref{IijDR}) does not depend on the Hadamard
regularization scales $s_1$ and $s_2$, which cancel out from the two
terms in the RHS. Therefore it is possible to re-express these two terms
(separately) by means of the constants $r'_1$ and $r'_2$ instead of
$s_1$ and $s_2$, where $r'_1$, $r'_2$ are the two fiducial scales
entering the Hadamard-regularization result~(\ref{IijH}). This
replacement being made the pHS term in Eq.~(\ref{IijDR}) is exactly the
same as the one in Eq.~(\ref{IijH}). At this stage all elements are in
place to prove the following theorem~\cite{BDEI04, BDEI05dr}.

\begin{theorem}
The DR quadrupole moment~(\ref{IijDR}) is physically equivalent to the
Hadamard-regularized one (end result of Refs.~\cite{BIJ02, BI04mult}),
in the sense that
\begin{equation}\label{shift}
\mathrm{I}_{ij}^{(\mathrm{HR})} = \lim_{\varepsilon\rightarrow 0}\left[
\mathrm{I}_{ij}^{(\mathrm{DR})} +
\delta_{\bm{\xi}}\mathrm{I}_{ij}\right],
\end{equation}
where $\delta_{\bm{\xi}}\mathrm{I}_{ij}$ denotes the effect of the same
shifts as determined in Theorems~\ref{th1}--\ref{th2}, if and only if
the HR ambiguity parameters $\xi$, $\kappa$ and $\zeta$ take the unique
values
\begin{equation}\label{resdimreg}
\xi=-\frac{9871}{9240},\qquad
\kappa=0,\qquad\zeta=-\frac{7}{33}.
\end{equation}
Moreover, the poles $1/\varepsilon$ separately present in the two terms
in the brackets of~(\ref{shift}) cancel out, so that the physical
(renormalized or ``dressed'') DR quadrupole moment is finite and given
by the limit when $\varepsilon\rightarrow 0$ as shown in
Eq.~(\ref{shift}).
\label{th3}
\end{theorem}

\noindent
This theorem finally provides an unambiguous determination of the 3PN
radiation field by dimensional regularization.

It should be emphasized that though the values~(\ref{resdimreg})
represent the end result of dimensional regularization, several
alternative calculations have provided a check, independent of
dimensional regularization, for all the parameters~(\ref{resdimreg}).
In~\cite{BI04mult} we computed the 3PN binary's \textit{mass dipole
moment} $\mathrm{I}_{i}$ using Hadamard's regularization, and identified
$\mathrm{I}_{i}$ with the 3PN \textit{center of mass vector position}
$\mathrm{G}_{i}$, already known as a conserved integral associated with
the Poincar\'e invariance of the 3PN equations of motion in harmonic
coordinates~\cite{ABF01}. This yields $\xi + \kappa = - 9871/9240$ in
agreement with Eq.~(\ref{resdimreg}). Next, we
considered~\cite{BDI04zeta} the limiting physical situation where the
mass of one of the particles is exactly zero (say, $m_2=0$), and the
other particle moves with uniform velocity. Technically, the 3PN
quadrupole moment of a \textit{boosted} Schwarzschild black hole is
computed and compared with the result for $\mathrm{I}_{ij}$ in the limit
$m_2=0$. The result is $\zeta = - 7/33$, and represents a direct
verification of the global Poincar\'e invariance of the wave generation
formalism. Finally, $\kappa = 0$ is proven~\cite{BDEI05dr} by showing
that there are no dangerously divergent ``diagrams'' corresponding to
non-zero $\kappa$-values, where a diagram is meant here in the sense of
Ref.~\cite{Dgef96}. All these verifications confirm the validity of
dimensional regularization for describing the dynamics of systems of
compact bodies.

\section{Conclusion}\label{secV}
The determination of the values~(\ref{lambda}) and~(\ref{resdimreg})
completes the problem of the general relativistic prediction for the
templates of inspiralling compact binaries up to 3PN order, and actually
up to 3.5PN order as the corresponding ``tail terms'' composing this
order have already been determined~\cite{B98tail}. The relevant
combination of the parameters~(\ref{resdimreg}) entering the 3PN energy
flux in the case of circular orbits, namely $\theta$ \cite{BIJ02}, is
now fixed to
\begin{equation}
\theta\equiv\xi+2\kappa+\zeta=-\frac{11831}{9240}\,.
\label{theta}\end{equation}
The orbital phase of compact binaries, in the adiabatic inspiral regime
(\textit{i.e.}, evolving by radiation reaction), involves at 3PN order a
linear combination of $\theta$ and of the equation-of-motion related
parameter $\lambda$ \cite{BFIJ02}, which is determined as
\begin{equation}
\hat{\theta}\equiv \theta-\frac{7}{3}\lambda=\frac{1039}{4620}\,.
\label{thetahat}\end{equation}
The parameter $\lambda$ appears here because the orbital phase follows
from energy balance between the total radiated energy flux and the
decrease of orbital center-of-mass energy which is computed from the
equations of motion.

The practical implementation of the theoretical templates in the data
analysis of detectors follows the standard matched filtering technique.
The raw output of the detector $o(t)$ consists of the superposition of
the real gravitational wave signal $h_{\rm real}(t)$ and of noise
$n(t)$. The noise is assumed to be a stationary Gaussian random
variable, with zero expectation value, and with (supposedly known)
frequency-dependent power spectral density $S_n(\omega)$. The
experimenters construct the correlation between $o(t)$ and a filter
$q(t)$, {\it i.e.}
\begin{equation}\label{ct}
c(t) = \int^{+\infty}_{-\infty} dt' o (t') q(t+t'),
\end{equation}
and divide $c(t)$ by the square root of its variance, or correlation
noise. The expectation value of this ratio defines the filtered
signal-to-noise ratio (SNR). Looking for the useful signal $h_{\rm
real}(t)$ in the detector's output $o(t)$, the experimenters adopt for
the filter
\begin{equation}\label{qtilde}
\tilde q (\omega) = {\tilde h (\omega)\over S_n (\omega)},
\end{equation}
where ${\tilde q} (\omega)$ and ${\tilde h} (\omega)$ are the Fourier
transforms of $q(t)$ and of the {\it theoretically computed} template
$h(t)$. By the matched filtering theorem, the filter~(\ref{qtilde})
maximizes the SNR if $h(t)=h_{\rm real}(t)$. The maximum SNR is then the
best achievable with a linear filter. In practice, because of systematic
errors in the theoretical modelling, the template $h(t)$ will not
exactly match the real signal $h_{\rm real} (t)$, but if the template is
to constitute a realistic representation of nature the errors will be
small. This is of course the motivation for computing high order
post-Newtonian templates, in order to reduce as much as possible the
systematic errors due to the unknown post-Newtonian remainder. The fact
that the numerical value of the parameter~(\ref{thetahat}) is quite
small, $\hat{\theta}\simeq 0.22489$, indicates, following
measurement-accuracy analyses~\cite{BCV03a, BCV03b}, that the 3PN (or
better 3.5PN) templates should constitute an excellent approximation for
the analysis of gravitational wave signals from inspiralling compact
binaries.

\begin{theacknowledgments}
It is a pleasure to thank Thibault Damour, Gilles Esposito-Far\`ese and
Bala Iyer for the exciting collaborations~\cite{BDE04, BDEI05dr}
summarized here, and to warmly acknowledge Jean-Michel Alimi for
organizing a very interesting conference.
\end{theacknowledgments}

\appendix 
\section{Useful formulas in \lowercase{$d$} spatial dimensions}\label{secA}
This Appendix is intended to provide a compendium of (mostly well-known)
formulas for working in a space with $d$ dimensions. As usual, though we
shall motivate some formulas below by writing some intermediate
expressions which make complete sense only when $d$ is a strictly
positive integer, our final formulas are to be interpreted, by complex
analytic continuation, for a general complex dimension,
$d\in\mathbb{C}$. Actually one of the main sources of the power of
dimensional regularization~\cite{tHooft, Bollini, Breitenlohner,
Collins} is its ability to prove many results by invoking complex
analytic continuation in $d$.

We discuss first the volume of the sphere having $d-1$ dimensions,
\textit{i.e.} embedded into Euclidean $d$-dimensional space. We separate out the
infinitesimal volume element in $d$ dimensions into radial and angular
parts,
\begin{equation}
d^d\mathbf{x}=r^{d-1}dr\,d\Omega_{d-1},
\label{ddx}
\end{equation}
where $r=\vert\mathbf{x}\vert$ denotes the radial variable
(\textit{i.e.}, the Euclidean norm of $\mathbf{x}\in\mathbb{R}^d)$ and
$d\Omega_{d-1}$ is the infinitesimal solid angle sustained by the unit
sphere with $d-1$ dimensional surface. To compute the volume of the
sphere, $\Omega_{d-1}=\int d\Omega_{d-1}$, one notices that the
following $d$-dimensional integral can be computed both in Cartesian
coordinates, where it reduces simply to a Gaussian integral, and also,
using (\ref{ddx}), in spherical coordinates:
\begin{equation}
\int d^d\mathbf{x}\,e^{-r^2}=\left(\int dx\,
e^{-x^2}\right)^d=\pi^{\frac{d}{2}}=\Omega_{d-1}\int_0^{+\infty}dr\,
r^{d-1}e^{-r^2}=
\frac{1}{2}\Omega_{d-1}\Gamma\left(\frac{d}{2}\right),
\label{intd}
\end{equation}
where $\Gamma$ in the last equation denotes the Eulerian function. This
leads to the well known result
\begin{equation}
\Omega_{d-1}=\frac{2\pi^{\frac{d}{2}}}{
\Gamma\left(\frac{d}{2}\right)}.
\label{Omegad1}
\end{equation}
For instance one recovers the standard results $\Omega_2=4\pi$ and
$\Omega_1=2\pi$, but also $\Omega_0=2$, which can be interpreted by
remarking that the sphere with 0 dimension is actually made of two
points. If we parametrize the sphere $\Omega_{d-1}$ in $d-1$ dimensions
by means of $d-1$ spherical coordinates $\theta_{d-1}$, $\theta_{d-2}$,
$\cdots$, which are such that the sphere $\Omega_{d-2}$ in $d-2$
dimensions is then parametrized by $\theta_{d-2}$, $\theta_{d-3}$,
$\cdots$, and so on for the lower-dimensional spheres, then we find that
the differential volume elements on each of the successive spheres obey
the recursive relation
\begin{equation}
d\Omega_{d-1}=\left(\sin\theta_{d-1}\right)^{d-2}
d\theta_{d-1}d\Omega_{d-2}.
\label{recursiondOmega}
\end{equation}
Note that this implies
\begin{equation}
\frac{\Omega_{d-1}}{\Omega_{d-2}}=\int_0^\pi
d\theta_{d-1}\left(\sin\theta_{d-1}\right)^{d-2}=
\int_{-1}^{+1}dx\left(1-x^2\right)^{\frac{d-3}{2}},
\label{ratioOmega}
\end{equation}
which can also be checked directly by using the explicit expression
(\ref{Omegad1}).

Next we consider the Dirac delta-function $\delta^{(d)}(\mathbf{x})$
in $d$ dimensions, which is formally defined, as in ordinary
distribution theory \cite{Schwartz}, by the following linear form
acting on the set $\mathcal{D}$ of smooth functions $\in
C^\infty(\mathbb{R}^d)$ with compact support:
$\forall\varphi\in\mathcal{D}$,
\begin{equation}
<\delta^{(d)},\varphi>\equiv\int
d^d\mathbf{x}\,\delta^{(d)}(\mathbf{x})\varphi(
\mathbf{x})=\varphi(\mathbf{0}),
\label{DiracBetter}
\end{equation}
where the brackets refer to the action of a distribution on
$\varphi\in\mathcal{D}$. Let us now check that the function defined by
\begin{subequations}\begin{eqnarray}
u&=&\tilde{k}\,r^{2-d},\\
\tilde{k}&=&\frac{\Gamma
\left(\frac{d-2}{2}\right)}{\pi^{\frac{d-2}{2}}},
\label{u}
\end{eqnarray}\end{subequations}
where $r$ is the radial coordinate in $d$ dimensions, such that
$r^2=\sum_{i=1}^d (x^i)^2$, is the Green's function of the Poisson
operator, namely that it obeys the distributional equation
\begin{equation}
\Delta u=-4\pi\delta^{(d)}(\mathbf{x}).
\label{Deltau}
\end{equation}
For any $\alpha\in\mathbb{C}$ we have $\Delta
r^\alpha=\alpha(\alpha+d-2)\,r^{\alpha-2}$, thus we see that $\Delta
u=0$ in the sense of functions. Let us formally compute its value in the
sense of distributions in $\mathbf{x}$-space.\footnote{The usual
verification of~(\ref{Deltau}) is done in Fourier space.} We apply the
distribution $\Delta u$ on some test function $\varphi\in\mathcal{D}$,
use the definition of the distributional derivative to shift the Laplace
operator from $u$ to $\varphi$, compute the value of the $d$-dimensional
integral by removing a ball of small radius $s$ surrounding the origin,
say $B(s)$, apply the fact that $\Delta u=0$ in the exterior of $B(s)$,
use the Gauss theorem to transform the result into a surface integral,
and finally compute that integral by inserting the Taylor expansion of
$\varphi$ around the origin. The proof of Eq.~(\ref{Deltau}) is thus
summarized in the following steps:
\begin{eqnarray}
<\Delta u,\varphi>&=&<u,\Delta\varphi>\nonumber\\
&=&\lim_{s\rightarrow
0}\int_{\mathbb{R}^d\setminus
B(s)}d^d\mathbf{x}\,u\Delta\varphi\nonumber\\
&=&\lim_{s\rightarrow
0}\int_{\mathbb{R}^d\setminus
B(s)}d^d\mathbf{x}\,\partial_i\left[u\partial_i\varphi-\partial_iu
\,\varphi\right] \nonumber\\
&=&\lim_{s\rightarrow 0}\int
s^{d-1}d\Omega_{d-1}(-n_i)
\left[u\partial_i\varphi-\partial_iu\,\varphi\right] \nonumber\\
&=&\lim_{s\rightarrow 0}\int
s^{d-1}d\Omega_{d-1}(-n_i)\left[-\tilde{k}\,(2-d)s^{1-d}n_i\varphi(
\mathbf{0})\right] \nonumber\\
&=&\Omega_{d-1}\tilde{k}\,(2-d)\varphi(\mathbf{0}) \nonumber\\
&=&-4\pi\varphi(\mathbf{0}).
\label{Deltauphi}
\end{eqnarray}
In the last step we used the relation between $\tilde{k}$ and the volume
of the sphere, which is
\begin{equation}
\tilde{k}\,\Omega_{d-1}=\frac{4\pi}{d-2}.
\label{ktildeBis}
\end{equation}
From $u=\tilde{k}\,r^{2-d}$ one can next find the solution $v$
satisfying the equation $\Delta v=u$ (in a distributional sense), namely
\begin{equation}
v=\frac{\tilde{k}\,r^{4-d}}{2(4-d)}.
\label{v}
\end{equation}
From (\ref{v}) we can then define a whole ``hierarchy'' of higher-order
functions $w$, $\cdots$ satisfying the Poisson equations $\Delta w=v$,
$\cdots$ in a distributional sense.

However, the latter hierarchy of functions $u$, $v$, $\cdots$ is better
displayed using some different, more systematic notation. This leads to
the famous Riesz kernels, here denoted $\delta_\alpha^{(d)}$, in
$d$-dimensional Euclidean space~\cite{Riesz}.\footnote{Besides the
Euclidean kernels $\delta_\alpha^{(d)}$, we also have the Minkowski
kernels (denoted $Z_A^{(d)}$)~\cite{Riesz}, which are at the basis of
the Riesz analytic continuation method~\cite{D83houches}.} These kernels
depend on a complex parameter $\alpha\in\mathbb{C}$, and are defined by
\begin{subequations}
\begin{eqnarray}
\delta^{(d)}_\alpha\left(\mathbf{x}\right)&=&K_\alpha\,
r^{\alpha-d},\\
K_\alpha&=&\frac{\Gamma\left(\frac{d-\alpha}{2}\right)}
{2^\alpha\pi^{\frac{d}{2}}\Gamma\left(\frac{\alpha}{2}\right)}.
\label{deltaalpha}
\end{eqnarray}\end{subequations}
For any $\alpha\in\mathbb{C}$, and also for any $d\in\mathbb{C}$, the
Riesz kernels satisfy the recursive relations
\begin{equation}
\Delta\delta^{(d)}_{\alpha+2}=-\delta^{(d)}_\alpha.
\label{recursiondeltaalpha}
\end{equation}
Furthermore, they obey also an interesting convolution relation, which
reads simply, with the chosen normalization of the coefficients
$K_\alpha$, as
\begin{equation}
\delta_\alpha^{(d)}\ast\delta_\beta^{(d)}
=\delta_{\alpha+\beta}^{(d)}.
\label{convolutiondeltaalpha}
\end{equation}
When $\alpha=0$ we recover the Dirac distribution in $d$ dimensions,
$\delta_0^{(d)}=K_0\,r^{-d}=\delta^{(d)}$ (the coefficient vanishes in
this case, $K_0=0$), and we have $u=4\pi\,\delta_2^{(d)}$,
$v=-4\pi\,\delta_4^{(d)}$, $\cdots$.

The convolution relation (\ref{convolutiondeltaalpha}) is nothing but
an elegant formulation of the Riesz formula in $d$ dimensions. To
check it let us consider the Fourier transform of $r^\alpha$ in $d$
dimensions,
\begin{equation}
\widetilde{f}_\alpha(\mathbf{k})\equiv\int
d^d\mathbf{x}\,|\mathbf{x}|^\alpha e^{-i\mathbf{k}.\mathbf{x}}.
\end{equation}
Using (\ref{ddx}) we can rewrite it as
\begin{equation}
\widetilde{f}_\alpha(\mathbf{k})=\int_0^{+\infty} dr
r^{\alpha+d-1}\int d\Omega_{d-1}e^{-i\mathbf{k}.\mathbf{x}},
\label{radialang}\end{equation}
in which the angular integration can be performed as an application of
Eq.~(\ref{recursiondOmega}). This yields an expression depending on the
usual Bessel function,
\begin{equation}
\int d\Omega_{d-1}e^{-i\mathbf{k}.\mathbf{x}}=\Omega_{d-2}\int_0^\pi
d\theta_{d-1}\left(\sin\theta_{d-1}\right)^{d-2}e^{-i k\,r
\cos\theta_{d-1}}=\left(2\pi\right)^{\frac{d}{2}}
\left(k\,r\right)^{1-\frac{d}{2}}J_{\frac{d}{2}-1}(k\,r),
\label{OmegaBessel}
\end{equation}
where $k\equiv\vert\mathbf{k}\vert$, and where we adopt for the Bessel
function the defining expression
\begin{equation}
J_\nu
(z)=\frac{\left(\frac{z}{2}\right)^\nu}{\Gamma\left(\nu+\frac{1}{2}
\right)\Gamma\left(\frac{1}{2}\right)}\int_{-1}^1
dx\left(1-x^2\right)^{\nu-\frac{1}{2}}e^{-i z x}.\label{Bessel}
\end{equation}

The radial integration in Eq.~(\ref{radialang}) is then readily done
from using the previous expression, and we obtain
\begin{equation}
\widetilde{f}_\alpha(\mathbf{k})=2^{\alpha+d}\,\pi^{
\frac{d}{2}}\,\frac{\Gamma\left(\frac{\alpha+d}{2}
\right)}{\Gamma\left(-\frac{\alpha}{2}\right)}\,k^{-\alpha-d},
\label{Fourier}
\end{equation}
where the factor in front of the power $k^{-\alpha-d}$, say $A_\alpha$,
is checked from the Parseval theorem for the inverse Fourier transform,
which necessitates that $A_\alpha A_{-\alpha-d}=(2\pi)^d$. To obtain
Eq.~(\ref{Fourier}) we employ the integration formula
\begin{equation}
\int_0^{+\infty}dz\, z^\mu J_\nu (z)=2^\mu\,\frac{
\Gamma\left(\frac{1+\mu+\nu}{2}
\right)}{\Gamma\left(\frac{1-\mu+\nu}{2}\right)}.\label{intformula}
\end{equation}
Finally we can check the Riesz formula by going to the Fourier domain,
using the previous relations. The result,
\begin{equation}
\int d^d\mathbf{x}\,r_1^\alpha r_2^\beta=\pi^{\frac{d}{2}}
\frac{\Gamma\left(\frac{\alpha+d}{2}\right)
\Gamma\left(\frac{\beta+d}{2}\right)
\Gamma\left(-\frac{\alpha+\beta+d}{2}\right)}
{\Gamma\left(-\frac{\alpha}{2}\right)
\Gamma\left(-\frac{\beta}{2}\right)
\Gamma\left(\frac{\alpha+\beta+2d}{
2}\right)}\,r_{12}^{\alpha+\beta+d},
\label{Riesz}
\end{equation}
is equivalent to Eq.~(\ref{convolutiondeltaalpha}).



\bibliographystyle{aipproc}   

\bibliography{/home/blanchet/Articles/ListeRef/ListeRef.bib}



\end{document}